\documentclass{aastex62}
\def\PLUTO{{\sc pluto}}
\def\CHOMBO{{\sc chombo}}

\shorttitle{On the physical foundations of equipartition}

\shortauthors{Uro{\v s}evi{\'c} et al.}

\begin{document}

\title{On the foundation of equipartition in supernova remnants}

\correspondingauthor{Dejan Uro\v sevi\'c}
\email{dejanu@matf.bg.ac.rs}

\author[0000-0003-0665-0939]{Dejan Uro\v sevi\'c}
\affiliation{Department of Astronomy \\
Faculty of Mathematics \\
University of Belgrade
Studentski trg 16, 11000 Belgrade, Serbia}
\affiliation{Isaac Newton Institute of Chile\\
Yugoslavia Branch}

\author[0000-0001-7633-8110]{Marko Z. Pavlovi\'c}
\affiliation{Department of Astronomy \\
Faculty of Mathematics \\
University of Belgrade
Studentski trg 16, 11000 Belgrade, Serbia}

\author[0000-0002-8036-4132]{Bojan Arbutina}
\affiliation{Department of Astronomy \\
Faculty of Mathematics \\
University of Belgrade
Studentski trg 16, 11000 Belgrade, Serbia}

%\author{D. Uro{\v s}evi{\'c}\altaffilmark{1,2}, M. Z. Pavlovi{\' c},
%\altaffilmark{1} and B. Arbutina\altaffilmark{1}}

%\altaffiltext{1}{Department of Astronomy, Faculty of Mathematics,
%University of Belgrade, Studentski trg 16, 11000 Belgrade, Serbia;
%dejanu@math.rs, arbo@math.rs}

%\altaffiltext{2}{Isaac Newton Institute of Chile, Yugoslavia Branch}

%\keywords{galaxies: individual (M82) --- ISM:  supernova remnants --- methods: statistical --- radio continuum: ISM}

\begin{abstract}
A widely accepted paradigm is that equipartition (eqp) between the energy density of cosmic rays (CRs) and the energy density of the magnetic field cannot be sustained in supernova remnants (SNRs). However, our 3D hydrodynamic supercomputer simulations, coupled with a non-linear diffusive shock acceleration model (NLDSA), provide evidence that eqp may be established at the end of the Sedov phase of evolution in which most SNRs spend the longest portions of their lives. We introduce term "constant partition" for any constant ratio between the CR energy density and the energy density of magnetic field in an SNR, while term "equipartition" should be reserved for the case of approximately the same values of the energy density (also, it is {constant partition} in the order of magnitude) of ultrarelativistic electrons only (or CRs in total) and the energy density of magnetic field. Our simulations suggest that this approximate constant partition does exist in all but the youngest SNRs. We speculate that since evolved SNRs at the end of the Sedov phase of evolution can reach eqp between CRs and magnetic fields, they may be responsible for initializing this type of eqp in the interstellar medium. Additionally, we show that eqp between the electron component of CRs and magnetic field may be used for calculating the magnetic field strength directly from observations of synchrotron emission from SNRs. The values of magnetic field strengths in SNRs given here are approximately 2.5 times lower than values calculated by \citet{arbo12, arbo13}.
\end{abstract}


\section{Introduction}

The importance of magnetic fields for the overall star-formation rate {\citep{li17}}, the gas dynamics of molecular clouds and the density and propagation of CRs is recognized widely \citep[for extensive reviews see][]{beck16, han17}. However, determining magnetic fields in the interstellar medium (ISM)
is a demanding task. There are three standard methods for determining the magnetic field strength in ISM: Zeeman splitting, rotation measure and equipartition calculation.
Zeeman splitting is often applied for very dense ISM, e.g. in molecular cloud cores.
By using Faraday rotation we can determine parallel component of the magnetic field in the direction of radio emission propagation. In this paper, we emphasize the significance of equipartition calculation for SNRs where strong magnetic field amplification provides conditions for global equipartition in ISM.

Equipartition (eqp) between the energy density of cosmic rays (CRs) and the energy density of magnetic fields is a common starting assumption in one of the methods for determining magnetic field strength \citep[see][]{pach70, bic05, arbo12, arbo13}. This eqp assumption is necessary for the calculation of magnetic field strengths by using the expression for the total energy in a synchrotron source. The expression for total energy density has two unknown variables (energy density of CRs ${\varepsilon_{\rm CR}}$ and the energy density of magnetic field ${\varepsilon_B}$). The expression which is necessary for solving the "equipartition problem" is the equation for synchrotron emissivity. The second expression -- a relation between ${\varepsilon_{\rm CR}}$ and ${\varepsilon_B}$ is obtained by differentiation of the equation for the total energy with respect to the independent variable $B$. With this we can calculate value for magnetic field strength for a minimum of total energy in the system. The eqp assumption is practically equivalent to the minimum energy requirement \citep[e.g. see][]{arbo12}.

In this paper, we introduce the concept of "partition" between CR and magnetic field energy densities in supernova remnants (SNRs) as a starting assumption, where the term equipartition describes an approximate equality (within an an order of magnitude) between the energy density of
ultra-relativistic electrons only (or all CRs) and magnetic fields associated with an SNR in which they produce the radio synchrotron emission. The electron eqp assumption leads us to a similar expression for eqp magnetic field as presented in \citet{arbo12, arbo13}. The calculated values of eqp magnetic fields in this paper are obtained for observed Galactic SNRs and they have lower values than the ones given in \citet{arbo12, arbo13}.

To provide evidence for the physical foundation of eqp (or constant partition) in SNRs, we carried out a set of hydrodynamic simulations of the SNR evolution coupled with a non-linear diffuse shock acceleration (NLDSA) model. In these simulations, the ratios between the energy density of CRs (or CR electrons) and the energy density of magnetic fields ${\varepsilon_{\rm CR}}$/${\varepsilon_B}$ (${\varepsilon_{\rm e}}$/${\varepsilon_B}$) are calculated through evolution of an SNR. If this ratio is approximately constant during the evolution in some period of time, the constant partition (or eqp) assumption should be physically justified in that period of evolution.

\section{Equipartition and constant partition}

Eqp in the interstellar medium (ISM) is assumed to be the result of redistribution of gravitational energy that formed the stars initially and converted into heat, radiation, etc. The "ingredients" of ISM (the electromagnetic radiation field, the random motion of gas, the magnetic field, and CRs) have similar energy densities, and they are all of the order of 1 eV/cm$^3$ in the vicinity of Sun \citep[see e.g.][]{long94,leq05,ptuskin07}. It was suggested by \citet{dju90}  that eqp between energy densities of magnetic fields and CRs holds in four nearby spiral galaxies and the values for these two energy density components are within one order of magnitude.

Relying on gamma-ray observations, \citet{yoast16} claim that the equipartition argument is frequently invalid in the  central molecular zones of nearby starburst galaxies M82, NGC 253 and Arp 220. However, their modeling also gives the magnetic field and the CR energy density approximately within one order of magnitude \citep[see Table 2 in][]{yoast16}, except in the case of Arp 220.

In this paper we use the 3D hydrodynamic simulations of SNR evolution in which we incorporate the production of high-energy relativistic particles (this production is based on the NLDSA model) and create the CR spectra. The produced electron spectra and amplified magnetic field values by non-linear effects are used for calculating radio synchrotron emissivity. Detailed explanation of the simulations and the model of particle acceleration used here can be found in \citet{pav17}. Our model combines the full 3D hydrodynamics, NLDSA and magnetic field amplification (MFA), allowing a self-consistent calculation of the CR spectra and the resulting continuum radiation. This emission can be fitted to present-day observations \citep[already well proven approach; see for example][although both based on 1D, spherically symmetric hydrodynamics]{bif06, lee12} and used to predict future or even past evolution. {3D simulations enable appropriate treatment of the structure in the mixing region between the forward and reverse shock, out of which the radio continuum emission mainly originates. Also, any deeper morphological studies of SNRs require a 3D
(or at least 2D) modeling \citep[see][and references therein]{blondin01, ferrand10}.}

We modeled the dynamical evolution of SNRs by numerically solving the time-dependent Euler partial differential equations (PDEs) of fluid dynamics, also known as hyperbolic conservation laws. Our simulations are performed by using the Godunov-type code for astrophysical gasdynamics {\PLUTO} \citep[Version 4.2;][]{mig07, mig12} with adaptive mesh refinement (AMR) implementation, based on the {\CHOMBO}\footnote{https://commons.lbl.gov/display/chombo} library. We improved {\PLUTO} code to model the hydrodynamic evolution of SNRs, self-consistently coupled with particle acceleration and the back-reaction of accelerated CRs in the amplified magnetic field. NLDSA calculation is done on the basis of the semi-analytical model of \citet{blasi05}, relying on the thermal leakage injection and containing most of the essential nonlinear effects we need. We adopted hydrodynamic equations to use the space and time-dependent adiabatic index $\gamma_{\rm eff}$, as a well established approach \citep[][first in 1D and the other two in 3D]{ellis04, ferrand10, orlando12}, also implemented and extensively tested in state-of-art AREPO code for cosmological simulations \citep{pfro17}. In terms of amplifying the magnetic field, for the first time in large scale SNR simulations, we include both resonant and non-resonant modes, by implementing models obtained from the first-principles, particle-in-cell (PIC) simulations \citep{capri14} and non-linear magnetohydrodynamic (MHD) simulations of CR-excited turbulence \citep{bell04}.

Throughout the SNR evolution, two different types of streaming instabilities are responsible for MFA \citep{amato09}. The non-resonant modes
are relevant mostly in the free expansion and early Sedov phase, while resonant waves dominate in a later SNR evolution. If non-resonant modes dominate, the amplified magnetic field saturates to a value
$B^2/8\pi \sim \frac{1}{2} \frac{\upsilon_{\rm s}}{c} \sigma_{\rm cr} \rho_0 \upsilon_{\rm s}^2$ \citep{bell04}, while \citet{capri14} showed that for resonant modes $B^2/B_0^2 \approx 3 \sigma_{\rm cr} M_{\rm A}$ is valid for MFA. We can then obtain the ratio between energy densities of non-resonantly and resonantly amplified magnetic fields:

\begin{equation}
\lambda \approx \frac{1}{3} \frac{\upsilon_{\rm s}}{c} M_{\rm A},
\end{equation}

\noindent where $B$ represents the amplified field, $B_0$ the ambient magnetic field strength, $\upsilon_{\rm s}$ the shock velocity, $\sigma_{\rm cr}$ the fraction of the shock energy in cosmic rays, $M_{\rm A}$ the Alfv\'enic Mach number and $\rho_0$ the ambient medium density. Therefore, we introduce correction $(1+\lambda)$ to the original relation
for resonant MFA \citep{caprioli09} in order to account for both resonant and non-resonant streaming instabilities:

 \begin{equation}
\label{eq:Pw-res}
\frac{P_{\mathrm{w},p}}{\rho_{0}\upsilon_{\rm s}^{2}}=\frac{1-\zeta}{4M_{A,0}} U_p^{-3/2}(1-U_p^{2}) (1+\lambda).
\end{equation}

\noindent Here, $P_{\mathrm{w},p}$ denotes precursor magnetic pressure of Alfv\'en waves
at point "$p$" in the precursor, $U_p$ represents the dimensionless fluid velocity $\upsilon_p/\upsilon_{\rm s}$ and $\zeta$ is the Alfv\'en wave dissipation parameter \citep[see][]{blasi02}. The ratio $\lambda \to 0$ as SNR approaches the later Sedov phase and therefore,
Equation~\ref{eq:Pw-res} reduces to the original equation where resonant MFA dominates.

Globally, our simulations are hydrodynamic (HD). Hydrodynamics is solved with {\PLUTO} code, while NLDSA module accounts for magnetic field and its dynamical effects. Throughout our modeling, we do not activate the MHD module of {\PLUTO} because it is, in any case, powerless in describing the generation of magnetic turbulence by the CRs upstream of the shock and corresponding MFA. Such an amplified magnetic field is dominant compared to the field compressed only due to fluid compression (which could be addressed with regular MHD approach), especially for young SNRs where non-linearity is very pronounced. Our NLDSA module, running simultaneously with HD code, performs calculations of MFA, synthesizes the global synchrotron emission in this amplified field and also accounts for its impact on hydrodynamics. Blasi's model \citep{blasi02} solves the velocity  profile of the thermal fluid by involving 4 terms: dynamical pressure, thermal pressure, non-thermal pressure of CRs and waves pressure (magnetic pressure). So the final compressibility of the fluid, applied at the cells near forward shock in HD simulation, is pretty much affected by the magnetic field.

HD simulations describe the evolution of SNRs in good agreement with analytical models of \citet{tmc99}. A reasonable level of difference is quite expected  due to the back-reaction from the pressure of CRs on the flow. Moreover, \citet{pfro17} showed with AREPO that CR acceleration at blast waves does not significantly break the self-similarity of the Sedov solution. They also concluded that the resulting modifications can be well approximated by a suitably adjusted adiabatic index, as done in our approach.

Among other, the results of our simulations are the ratios between CR (electron) and magnetic field energy densities through SNR evolution. Our simulations follow SNR evolution starting from $\approx$ 30 yr after the SN explosion to the end of Sedov phase, where the adiabatic condition is completely applicable.

In all of our simulations, we use classical NLDSA model, predicting the concave CR spectrum between the injection momentum and the cutoff
momentum, although gamma-ray observations point to significantly steeper high-energy CR spectra \citep{capri12, kang13}. A possible explanation lies
in a very efficient MFA and very pronounced magnetic turbulence, causing strong Alfv\'en drift in the early phases of SNR evolution, but this should not have effect on our conclusions for the timescale considered. 

%As an illustration, we also run considerably shorter simulations with Alfv
%\'en drift included.

We expect that particle spectra (both proton and electron), obtained in the NLDSA model, should be a good representation of the real particle spectra in SNRs. Otherwise, it would not be in good agreement with the observed radiation properties. We carefully tuned a set of input parameters for our simulations in order to obtain the SNR flux densities which we know from radio observations. This set of input parameters is rather unique, i.e. a variation of each input parameter must be small if we want to obtain, as the primary goal, the observed flux densities. Further explanations about the choice of input parameter values are presented in the following sections of this paper.

\begin{figure}
\includegraphics[width=\columnwidth]{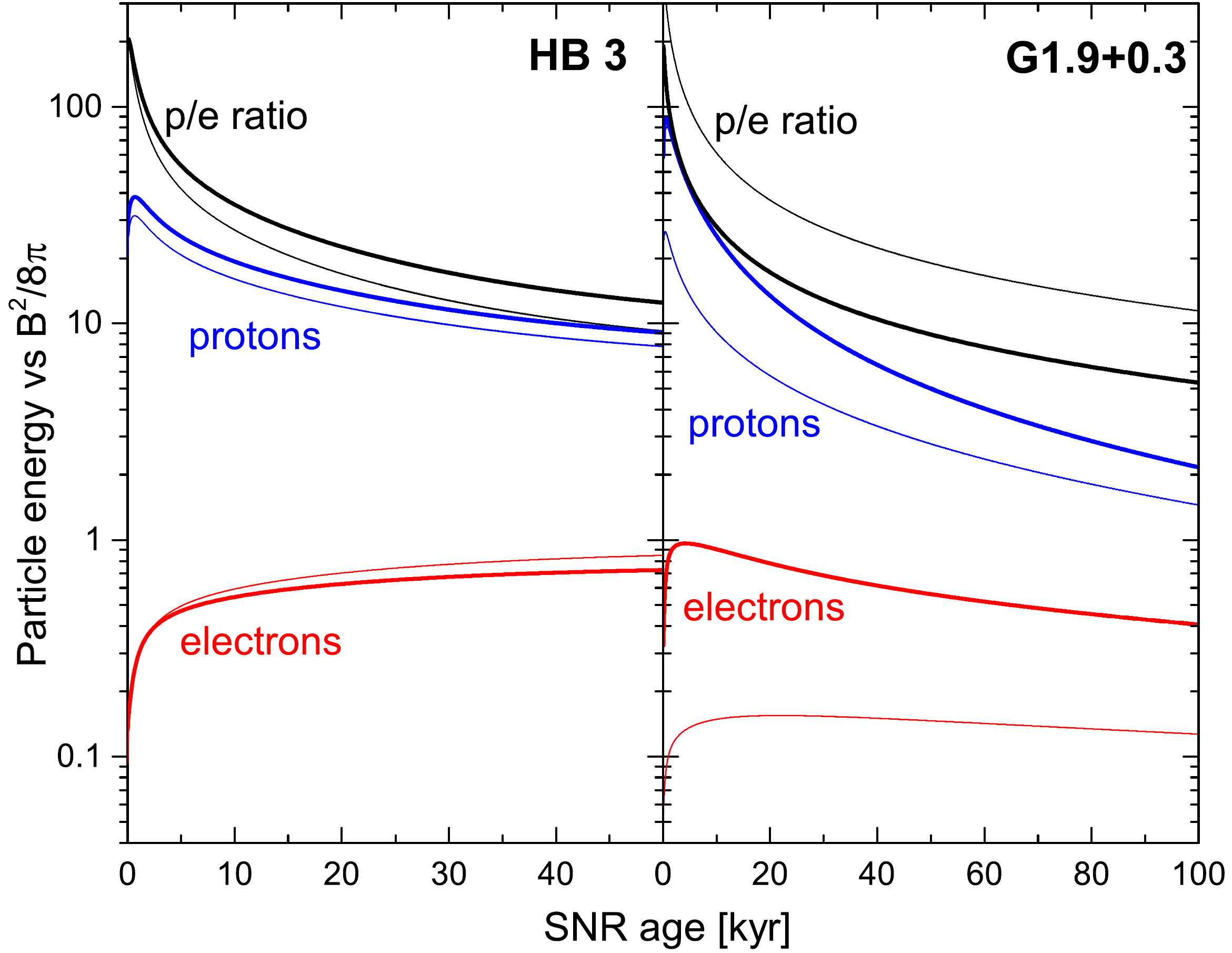}
\caption{Temporal evolutions of the ratios between CR proton (electron)
kinetic energy density and magnetic field energy density at the SNR shock front, with assumed $K_{\rm ep} = 0.01$, are represented with thick blue (red) lines. Simulation parameters are carefully tuned to reproduce current observational properties of the two particular SNRs, namely an evolved SNR -- HB 3 evolving in a dense medium and the youngest Galactic SNR G1.9+0.3 evolving in a rarefied medium. We also give the ratios between proton and electron kinetic energy densities during the lives of the SNRs (thick black line). The thin lines represent evolutions of the same ratios with assumed $K_{\rm ep} = 0.007$ for HB3 and $K_{\rm ep} = 0.004$ for G1.9+0.3.}
\end{figure}

According to the results of simulations presented in Figure 1, we can assume an approximate eqp between the ultrarelativistic electron and magnetic field energy densities. Although, we lack a physical justification for this eqp, this assumption should give more reliable magnetic field estimates from eqp i.e. minimum-energy calculation. From Equation 2 in \citet{arbo12} we obtain for energy density of electrons:

\begin{equation}
\varepsilon  =  K (m_{\mathrm{e}} c^2)^{2-\gamma}\frac{\Gamma
(\frac{3-\gamma}{2})\Gamma (\frac{\gamma -2}{2})
}{2\sqrt{\pi}(\gamma -1)},
\end{equation}

\noindent where $K$ is the constant in the power-law energy distributions for electrons and $\gamma$ is the energy index of ultrarelativistic electrons. If we want to find the minimum of total energy using the expression for the electron energy density (3), and following the same derivation as presented in \citet{arbo12}, we obtain a similar equation for the calculation of the magnetic field values as was shown in \citet[][specifically their Equation 12]{arbo12}. Our Equation for the magnetic field in the minimum total energy condition (sum of electron and magnetic field energies) gives $(1+\kappa)^{2/(\gamma +5)}$ times\footnote{$\kappa$ is defined in \citet{arbo12} and represents the energy ratio between ions and electrons.} lower values than in \citet{arbo12}, and has the following form:

\begin{eqnarray}
 B\ \mathrm{[Ga]} &\approx & \Big(6.286\cdot 10^{(9\gamma -79)/2}
\frac{\gamma +1}{\gamma -1}\frac{\Gamma
(\frac{3-\gamma}{2})\Gamma(\frac{\gamma -2}{2})\Gamma
(\frac{\gamma +7}{4})}{\Gamma (\frac{\gamma +5}{4})} (m_{\mathrm{e}}
c^2)^{2-\gamma}
 \cdot   \\
 &\cdot &  \frac{(2c_1)^{(1-\gamma)/2}}{c_5}  \frac{S_\nu \mathrm{[Jy]}}{f\ d
 \mathrm{[kpc]}\
 \theta
[\mathrm{arcmin}]^3} \nu[\mathrm{GHz}] ^{(\gamma
-1)/2}\Big)^{2/(\gamma +5)}, \nonumber
\end{eqnarray}

\noindent

\noindent where $m_{\mathrm{e}} c^2 \approx 8.187\cdot 10^{-7}$ ergs, $S_\nu$ is the flux density, $f$ is the volume filling factor, $d$ is the distance to the SNR, $\theta$ is the angular diameter, and $\nu$ is the frequency. We also have:
\begin{equation}
\varepsilon_B = \frac{\gamma +1}{4} \varepsilon_{\mathrm{e}}, \ \ \ \varepsilon_{\mathrm{min}}
= \frac{\gamma +5}{\gamma +1} \varepsilon_B,
\end{equation}

\noindent where $\varepsilon_{\mathrm{min}}$ is the minimum total energy density. These are the same expressions as presented in \citet{arbo12}.

The concept of constant partition has already been mentioned in \citet[][Equation 28]{arbo12}. If ${\varepsilon_B}/{\varepsilon_{\rm CR}}=\beta=\rm{const}$, then:

\begin{equation}
B^\prime=\Big(\frac{4\beta}{\gamma+1}\Big)^{2/(\gamma +5)}B,
\end{equation}

\noindent where $B^\prime$ is the recalculated field for $\beta$ = const, while $B$ is the field corresponding to the minimum energy.

\subsection{Examples: SNRs G1.9+0.3 and HB3}

We chose two SNRs in order to check the possibility of eqp (or constant partition). The first one is the remnant of SN Ia (G1.9+0.3) which evolves in a low-density ambient medium \citep[$\sim 0.02~\rm{cm}^{-3}$;][]{pav17}, and the second one is the remnant of the core-collapse SN (HB3, Figure 2) which evolves in a dense environment \citep[we use $\sim 0.5~\rm{cm^{-3}}$ as an intermediate value between values obtained in][]{gosa05, laze06}. {SNR G1.9+0.3 is a typical representative of a group of young remnants of {type Ia SN} explosions which are embedded in low density environment. On the other hand, SNR HB3 (G132.7+1.3) is a typical evolved, mixed-morphology remnant of core-collapse SN in interaction with dense molecular cloud environment. These SNRs are quite opposite of each other concerning the initial conditions for their entire evolution. Because of this we chose these two SNRs as typical representatives of their groups for detailed examination in this paper. In future, we plan to analyze all young SNRs: both Balmer dominated and oxygen rich, as well as all evolved SNRs, especially those embedded in dense environment, in order to examine eqp in SNRs as well as in the entire ISM.} As emphasized before, in order to calibrate our simulations we used data obtained by observations. As the reference radio data for SNR G1.9+0.3, we use observations made by \citet{green08} at 1.43 and 4.86 GHz with the VLA, which are respectively 
0.935 $\pm$ 0.047 Jy and 0.437 $\pm$ 0.022 Jy, and a radio spectral index of $\alpha \approx 0.6$. For HB3 we refer to the flux measurement of 29.4 $\pm$ 2.7 Jy and a spectral index of $\alpha=0.6$ at 1420 MHz \citep{kothes06}.

\begin{figure}
\includegraphics[width=\columnwidth]{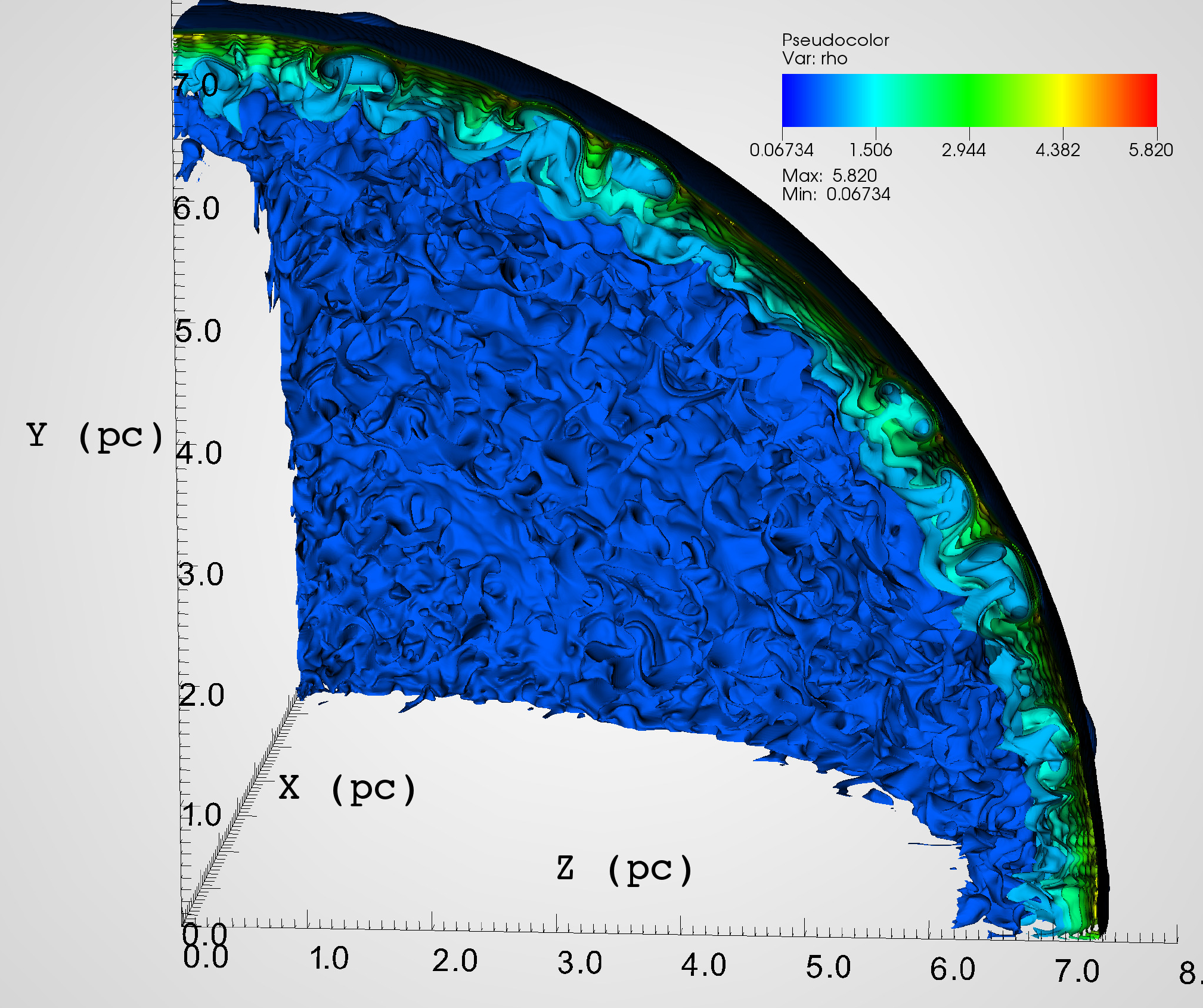}
\caption{3D rendering of the spatial distribution of the plasma number density for SNR HB3, on a linear scale, in units of $\rm{cm}^{-3}$ (see color table in the upper right-hand corner of the panel), after 2000 yr of evolution through homogeneous ISM with an ambient density of 0.5 $\rm{cm}^{-3}$. We applied the isosurface slicing operator to extract 10 density levels (linear scale) from the 3D HDF5 database \citep[credit by VisIt,][]{brugg12}.}
\end{figure}

For the injection of electrons, we use electron to proton number density ratio $K_{\rm ep} = 10^{-2}$, as observed in local fluxes of cosmic-ray protons and electrons around 10 GeV {\citep[also theoretically explained by][]{bell78}}. This may be a strong assumption, but is not in conflict with observations \citep{yuan12}. It may affect only the final value for electron to magnetic field energy density ratio but not its qualitative behavior during SNR lifetime, as explained above. Even more importantly, some multiwavelength observations of mainly young SNRs suggest $K_{\rm ep} = 10^{-3}$ or less \citep{volk05, morlino12}, while higher values around $10^{-2}$ seem to be characteristic of the later stages of the SNR evolution \citep[e.g.][]{sarb17}. Therefore, uncertainties of the $K_{\rm ep}$ value are probably less pronounced in later stages when our simulations give approximate eqp between CR electrons and magnetic field. {Additionally, in order to check validity of the adopted $K_{\rm ep} = 10^{-2}$, we performed a set of additional simulations with assumed smaller values of this parameter. The main {criteria} for validity was to reproduce the observed flux densities with lower assumed values of parameter $K_{\rm ep}$ for both SNRs. If we assume $K_{\rm ep} = 10^{-3}$, we cannot reproduce observed flux densities for both SNRs, {while keeping} other parameters in physically meaningful limits (see Figure 3). The CR injection parameter $\xi$ can typically be in the range 3.0$-$4.5 \citep{kosenko14}. Parameter $\zeta$ determines the amount of MHD waves dissipation through the heating of plasma and it can theoretically take values from $0$ to $1$ \citep{caprioli09}. However, we avoid 'extreme' values ($\zeta=0$ or $1$) in our simulations as intermediate cases seem more reasonable in reality \citep{kang13}.
We can obtain the observed flux density value by using $K_{\rm ep} = 0.004$ for G1.9+0.3 ({$\xi=3.34$, $\zeta=0.34$; see} Figure 3, right panel). By using $K_{\rm ep} = 0.007$ for HB3, the observed flux density cannot be {fully reproduced} ({$\xi=3.20$, $\zeta=0.10$;} see Figure 3, left panel). Due to this we have indications that the assumption $K_{\rm ep} = 10^{-2}$ is reasonable and can account for the observed flux densities for both SNRs. {Also, we confirm the above mentioned \citep{volk05, morlino12, sarb17}, higher deviation of $K_{\rm ep}$ from the value $10^{-2}$ in case of young SNRs such as G1.9+0.3 than for an old SNR HB3.} To check {in} which way our simulations depend on the assumed value of parameter $K_{\rm ep}$, we {performed} simulations with assumed $K_{\rm ep} = 0.004$ for G1.9+0.3 and with $K_{\rm ep} = 0.007$ for HB3. The results are shown in Figure 1. Obviously, the presented ratios show that the variation in the assumed value of $K_{\rm ep}$ does not change trends and quantitative conclusions significantly. A somewhat larger difference (around 3.5 times smaller value) is obtained for the electron to magnetic field energy density ratios for G1.9+0.3 (Figure 1). On the other hand, this difference of 3.5 times can produce only $25\%$ smaller estimate for the eqp magnetic field values from the electron eqp method suggested in this paper. In view of the presented facts, we conclude that $K_{\rm ep} = 10^{-2}$ is a validly assumed value for our two representative SNRs, and this is the assumption underlying further analysis and conclusions herein.

\begin{figure}
\includegraphics[width=\columnwidth]{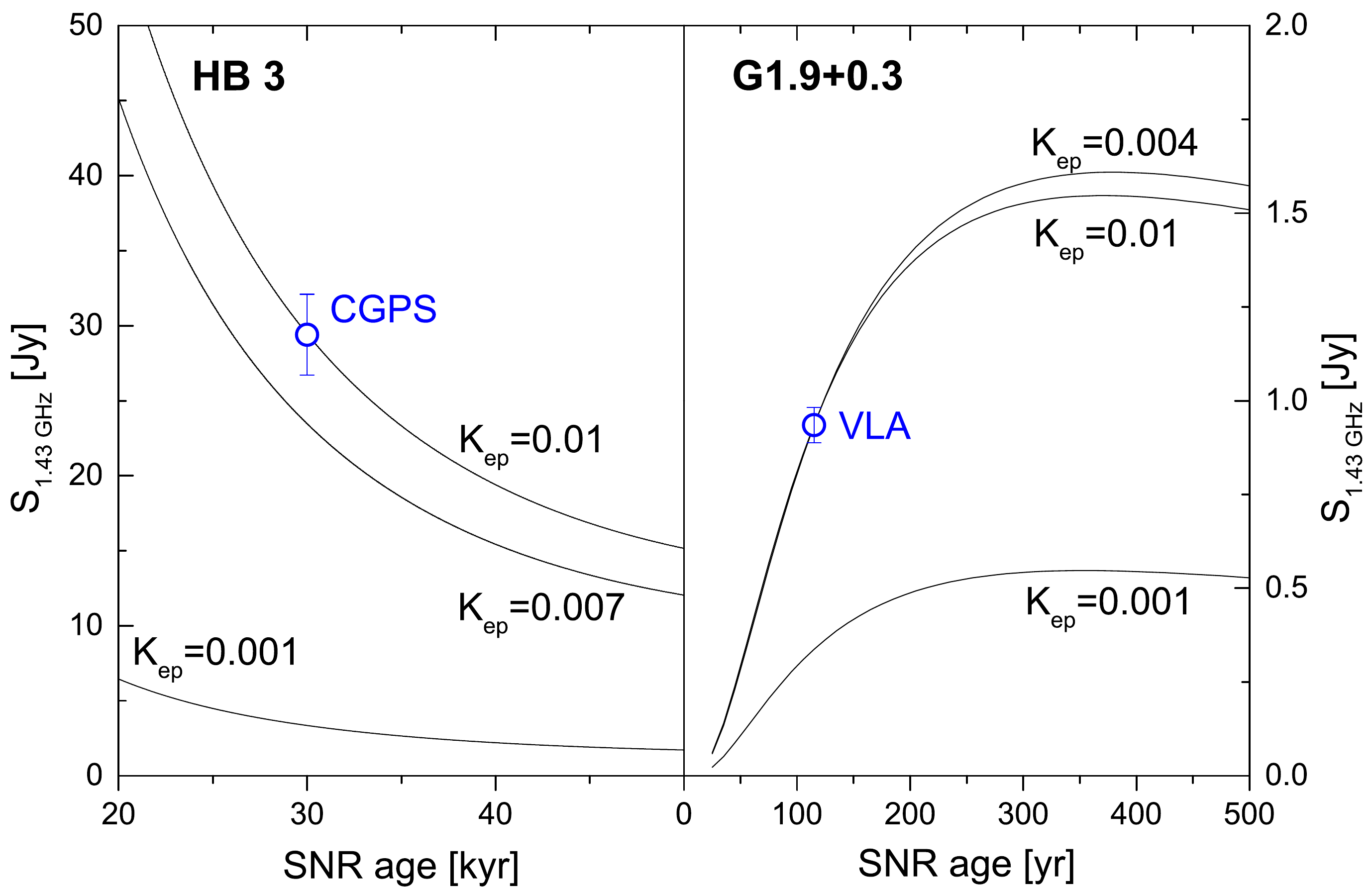}
\caption{The simulated flux density evolutions with different assumed values for the injected electron to proton number density ratio $K_{\rm ep}$. Open circles represent the observationally measured values of flux densities.}
\end{figure}

Our initial conditions were carefully tuned to reproduce G1.9+0.3 after around 120 yr of evolution with a  shock radius of 2 pc, an assumed location near a Galactic Center \citep{reynolds08}, and a shock velocity of 14\,000 km s$^{-1}$ \citep{borkow10}. We assumed an initial spherical remnant with a radius of 0.5 pc (corresponding to an initial age of $\approx$ 30 yr), ejecta mass $M_{\rm ej} = 1.4 M_{\sun}$, total explosion energy of $E_0=10^{51}$ erg, expanding through a homogeneous rarefied plasma with a hydrogen number density of $n_{\rm H}=0.02$ cm$^{-3}$ and temperature $T = 10^4$ K. We use typical NLDSA parameters, namely injection parameters $\xi=3.4$ and local wave dissipation $\zeta=0.7$ \citep[see e.g.][]{pav17}. The maximum number of AMR levels used in the 3D simulation with base grid $32^3$ gradually decreased from 10 (initially) to 4 (at the final time) following the SNR expansion. The effective mesh size varied from $32768^3$ initially to $512^3$ at the final time (100 kyr). After inspection of the results of simulation presented in Figure 1, we conclude that the assumption of constant partition is not justified for the first 90 kyr of SNR evolution, with the ratio between CR protons and magnetic field energy densities declining in value by an order of magnitude. However, this change is not huge for such a long period of SNR evolution. Furthermore, the ratio of CR electron to magnetic field energy densities we find to be more constant in a period of 90 kyr (from 10 to 100 kyr from explosion) of evolution for SNR G1.9+0.3 (see Figure 1). The approximate eqp (${\varepsilon_{\rm e}}$/${\varepsilon_B}\sim 0.5$, Figure 1) between ultrarelativistic electron and magnetic field energy densities is justified to within a variation of 20\% in 90 kyr of the SNR evolution. The magnetic field value for SNR G1.9+0.4 calculated in \citet{arbo12} is approximately 175 $\mu$G. The eqp value suggested in this paper, based on electron eqp is 75 $\mu$G. It is simply obtained when the value from \citet{arbo12} is lowered by $(1+\kappa)^{2/(\gamma +5)}$ times, where $\kappa$ is calculated for 100\% hydrogen plasma. On the other hand, this SNR is only 120 yr old and the eqp (or constant partition) is not a valid assumption for the calculation of magnetic field value in the present moment of evolution (see Figure 1) -- but after 10 kyr in future it will be. Directly from the simulations we obtain the magnetic field strength of 300 $\mu$G. An interesting fact to note is that both previous eqp values are approximately in the order of magnitude to the correct value obtained from simulation.

The non-thermal X-ray emission in the young SNR G1.9+0.3 is most likely synchrotron emission, and seems to be ideal for comparison with simulation in order to constrain the energy densities. In our approach, used for long-term simulations, we follow the  receipt for electron spectrum cutoff $\propto \exp\left[-(p/p_{\rm{e,max}})^2\right]$, as suggested by \citet{zirak07} for the loss-dominated case. We calculate the electron maximum momentum $p_{\rm{e,max}}$ in the Bohm diffusion regime, by using the approximate implicit expression, determined by \citet{morlino09}. However, synchrotron losses can be consistently taken into account only by supplementing the ordinary diffusive transport equation by a corresponding loss term \citep[as done, e.g. in][]{bif02,bif04}. Our approach overestimates X-ray emission for G1.9+0.3 and should be additionally tuned in high-energy part of the electron spectrum, in order to obtain satisfying model of synchrotron X-ray spectra which fits well with observations, as described in \citet{pav17}. After applying here appropriate model from \citet{pav17}, in order to fit currently observed X-ray spectrum for G1.9+0.3 \citep{reyn08,zog15,ahar17}, we obtain around 30\% lower electron to magnetic field energy density ratio ($\varepsilon_{\rm e}/\varepsilon_B \sim 0.34$ instead of previously obtained $0.48$, at $t=115~\rm{yr}$, for electron to proton number density ratio $K_{\rm ep}=0.01$), in comparison to values derived in Figure 1 of the present paper. Obtained X-ray spectrum is very similar to spectrum shown in Figure 10 of the paper \citet{pav17}. In case of $K_{\rm ep}=0.004$, this difference is around 25\% ($\varepsilon_{\rm e}/\varepsilon_B \sim 0.045$ instead of previously obtained $0.06$, at $t=115~\rm{yr}$). We believe that 30\% here represent maximum expected difference in electron energy density values, as in later stages of SNR evolution synchrotron losses became less pronounced due to less efficient acceleration and lower maximum energy for electrons. Therefore, we do not expect such a difference in evolved SNRs in the Sedov phase of evolution.

For SNR HB3 (G132.7+1.3), we assume a current age of around 30 kyr \citep{laze06} and a radius of $\approx$ 23 pc \citep[for a distance of 1.95 kpc,][]{zhou16}, total ejecta mass of $M_{\rm ej} = 10 M_{\sun}$ and an explosion energy of $E_0=2.5 \times 10^{51}$ erg. {Again, we assume a spherical remnant of an initial age of $\approx$ 30 yr with a radius of 0.5 pc.} NLDSA module ($\xi=3.4$, $\zeta=0.6$) gives the currently observed spectral characteristics, for an ambient hydrogen number density of $n_{\rm H}=0.5$ cm$^{-3}$. We set base 3D grid $32^3$, with maximum number of AMR levels going from 9 (initially) until 4 (at the final time), following the SNR expansion. The effective mesh size varied from $16384^3$ initially to $512^3$ at the final age of 50 kyr. Figure 2 shows simulated 3D spatial distribution of the plasma number density for HB3,
2000 yr after initial explosion. This simulation results in a constant partition (${\varepsilon_{\rm CR}}$/${\varepsilon_B}\sim 10$, Figure 1) with a slight declining trend of 50\% in 40 kyr (from 10 to 50 kyr from the initial explosion). By using Equation (6), with $\beta=1/12$, we obtain 20 $\mu$G. Additionally, the electron eqp concept is justified (${\varepsilon_{\rm e}}$/${\varepsilon_B}\sim 0.7$, Figure 1) with a rising of ratio of only 35\% during the 40 kyr of evolution.  The electron eqp calculated by using Equation (4), gives 17 $\mu$G (Table 1). {The magnetic field strength of 40 $\mu$G is obtained directly from the simulation. The result ${\varepsilon_{\rm e}}$/${\varepsilon_B}\sim 0.7$ with a rising ratio of 35\% in 40 kyr of evolution supports the conclusion that the electron eqp is a better approximation than "classical" (CR) eqp for all SNRs in which the parameter $\beta$ is unknown.} In other words, we can say that the CRs to magnetic field constant partition, if it exists, is approximately equivalent to the electron to magnetic field eqp in SNRs. The general conclusion is the following: the electron eqp calculation gives values for the magnetic fields which have relatively small deviations in comparison to the correct ones (obtained from simulation) through the specific period of SNR evolution -- with an accuracy of an order of magnitude. We calculated values for magnetic field strengths by using electron eqp (Equation 4) for the sample of Galactic SNRs from Pavlovi{\' c} et al. (2014) and showed them in Table 1.

Figure 1 also shows temporal evolution of the proton to electron energy density ratio during the expansion of SNRs. Interestingly, this ratio gradually decreases for roughly one order of magnitude or slightly more in the approximately same way for both SNRs, which evolve in widely different ambient media and have different initial explosion energies.

\section{Physical foundations of eqp}

The results of this paper also suggest a {possible} answer to the question as to how eqp between CR and magnetic field energy densities in the ISM can be achieved? As noted earlier the eqp between these two ingredients of the ISM is a widely accepted assumption. Our simulations show that SNR evolution in a low-density environment provides indication that CR protons and magnetic fields can be in eqp at the end of the Sedov phase of evolution. The CR proton to magnetic field energy density ratios tend toward approximate eqp (${\varepsilon_{\rm CR}}$/${\varepsilon_B}\rightarrow 2$, Figure 1, right panel). After the Sedov phase, an SNR should enter the radiative phase of evolution. The radiative shocks provide high compression ratios which depend on the square of shock Mach number. These high compression ratios would provide higher magnetic field energy densities but also higher CR energy densities. Due to this we can expect an approximate eqp until the end of the evolution of an SNR. DSA should not be significantly efficient on these low Mach number shocks (as noted by \citet{sre09}), in comparison to the processes based on high compression in downstream region. A detailed analysis of the radiative phase is beyond the scope of this paper, but will be the focus of our future work.

SNRs are strong synchrotron emitters from radio to X-rays. For this kind of radiation process the necessary ingredients are magnetic field and ultrarelativistic electrons. If these ingredients are located in some system and they have non-negligible energy densities (in comparison to each other), we can expect synchrotron radiation from the system. If ${\varepsilon_{\rm e}}\gg{\varepsilon_B}$ the synchrotron mechanism should be transformed into non-thermal bremsstrahlung. Observationally, SNRs appear as objects with harder spectra than expected for non-thermal bremsstrahlung emitters. For a population of ultrarelativistic electrons produced on the shock wave of an SNR, we expect an energy index of $\gamma=2$ \citep[see][and references therein]{uro14}. For an energy index of this value, the radio spectral index is $\alpha=0.5$. On the other hand, for the same population of electrons, the value of the spectral index corresponding to non-thermal bremsstrahlung radiation is $\alpha=1$ \citep[see e.g.][]{helder12}. Because of this we can conclude that SNRs are not strong non-thermal bremsstrahlung emitters\footnote{But for radio galaxies with steeper radio spectra and no detected polarization, the non-thermal bremsstrahlung can be a source of contamination to the pure synchrotron radiation.}. Furthermore, we can conclude that the energy density of magnetic field should not be negligible in comparison to the energy density of ultrarelativistic electrons for the efficient production of synchrotron radiation in SNRs. On the other hand, if ${\varepsilon_{\rm e}}\ll{\varepsilon_B}$, we can expect strong synchrotron radiation in $\gamma$ rays (as in the case of pulsars), but this kind of synchrotron radiation is not detected from SNRs. Analogously, the energy density of ultrarelativistic electrons should not be negligible in comparison to the energy density of magnetic field in SNRs. Our simulations, described in the previous Section, show that the ratios between these necessary synchrotron "ingredients" are between 0.1 and 0.9 for evolved SNRs (Figure 1) and that the assumption of electron eqp is correct -- meaning that these two energy pools should be similar for the production of synchrotron radiation to be efficient.

Additionally, there are some indications of eqp and constant partition between the energy densities of electrons and magnetic fields in some $\gamma$ TeV SNRs with the associated emission mainly based on the leptonic scenario -- these estimates of eqp depend on the distances to SNRs and should be accepted with caution (see \citet{yang14}, and references therein).

As mentioned in the previous Section, the eqp for an entire galaxy is expected \citep{dju90}, but for SNRs which represent the sources of CRs, the ratio ${\varepsilon_{\rm CR}}$/${\varepsilon_B}$ should be higher in an SNR in the starting phases of evolution than in an entire galaxy. Galactic CRs energy density should slightly drop due to the leakage of the highest energy CRs from a galaxy and the diffusion of CRs through the galaxy \citep{dju95, long94}. However, the highest energy CRs contain a negligible part of the total energy contingent of all CR particles. {Due to this we can expect that SNRs at the end of their evolution can provide the basis for eqp which then can be supported by processes related to turbulent motion throughout the entire galaxy.}

\section{Summary}

Based on the results obtained from 3D hydrodynamic supercomputer simulations of the SNR evolution, coupled with the production of CRs in a non-linear DSA model, we highlight the most important observations as follows:

\noindent i) Eqp is a justified assumption especially between the CR electrons and the magnetic fields in evolved SNRs in the Sedov phase of evolution (${\varepsilon_{\rm e}}/{\varepsilon_B} \simeq 0.1 - 0.9$, with averaged  value near 0.5).

\noindent ii) We provide evidence suggesting that electron eqp formulae should be used for the calculation of the magnetic field strengths in SNRs. The obtained values are approximately 2.5 times lower than those determined in earlier calculations.

\noindent iii) Evolved SNRs, especially those embedded in a rarefied ambient medium, at the end of the Sedov phase of evolution can reach eqp between CRs and magnetic fields similar to that in the ISM (${\varepsilon_{\rm CR}}/{\varepsilon_B} \sim 2$).

\section*{Acknowledgements}

This paper is a part of the project No. 176005 'Emission nebulae: structure
and evolution' supported by the Ministry of Education, Science and
Technological Development of the Republic of Serbia. Numerical
simulations were run on the PARADOX-IV supercomputing facility
at the Scientific Computing Laboratory of the Institute of Physics
Belgrade, supported in part by the Ministry of Education, Science
and Technological Development of the Republic of Serbia under
project No. ON171017. Simulations were also
run on cluster Jason, belonging to Automated Reasoning Group
(ARGO) based at the Department of Computer Science, Faculty
of Mathematics, University of Belgrade. The authors thank the anonymous referee for valuable comments that improved the quality of this paper and Dragana Momi\' c for English editing.
The authors thank Nebojsa Duric for his careful reading and editing of the manuscript, as well as Du{\v s}an Oni{\' c} for his valuable comments.  M.P. acknowledges Salvatore Orlando and Gilles Ferrand for useful discussions and advices. {\PLUTO} has been developed at the Turin Astronomical Observatory in collaboration with the Department of Physics of Turin University.

\software{PLUTO \citep[Version 4.2; ][]{mig07,mig12}, Visit \citep{brugg12}, Chombo \citep{adams13}}

\startlongtable
\begin{deluxetable}{cccc}
\tablecolumns{8}
\tablewidth{0pc}
\tablecaption{Calculated magnetic field strengths for the sample of Galactic SNRs from \citet{pav14}. $B_1$ is the field obtained by using simple approach from \citet{arbo12} and $B_2$ is obtained by setting $\kappa = 0$ i.e. by finding minimum-energy of the sum of magnetic field and CR electrons only.}
\tablehead{
\colhead{Name}    &   \colhead{Other name}   & $B_1$ [$\mu$Ga] &  $B_2$ [$\mu$Ga] \\
}
\startdata
G4.5+6.8	&	Kepler, SN1604, 3C358	&	365	&	172	\\
G11.2-0.3	&		&	326	&	141	\\
G18.1-0.1	&		&	235	&	82	\\
G21.8-0.6	&	Kes 69	&	173	&	60	\\
G23.3-0.3	&	W41	&	142	&	50	\\
G27.4+0.0	&	4C-04.71	&	197	&	101	\\
G33.6+0.1	&	Kes 79	&	207	&	72	\\
G35.6-0.4	&		&	157	&	54	\\
G46.8-0.3	&	HC30	&	126	&	44	\\
G53.6-2.2	&	3C400.2, NRAO 611	&	63	&	37	\\
G54.4-0.3	&	HC40	&	84	&	29	\\
G55.0+0.3	&		&	36	&	13	\\
G65.1+0.6	&		&	17	&	8	\\
G78.2+2.1	&	DR4, $\gamma$ Cygni SNR	&	159	&	55	\\
G84.2-0.8	&		&	108	&	38	\\
G93.7-0.2	&	CTB 104A, DA 551	&	51	&	24	\\
G96.0+2.0	&		&	32	&	11	\\
G108.2-0.6	&		&	41	&	14	\\
G109.1-1.0	&	CTB 109	&	107	&	37	\\
G111.7-2.1	&	Cassiopeia A, 3C461	&	1245	&	763	\\
G114.3+0.3	&		&	51	&	18	\\
G116.5+.1.1	&		&	48	&	17	\\
G116.9+0.2	&	CTB 1	&	55	&	24	\\
G119.5+10.2	&	CTA 1	&	38	&	17	\\
G120.1+1.4	&	Tycho, 3C10, SN1572	&	280	&	135	\\
G132.7+1.3	&	HB3	&	40	&	17	\\
G152.4-2.1	&		&	21	&	10	\\
G156.2+5.7	&		&	30	&	11	\\
G160.9+2.3	&	HB9	&	46	&	22	\\
G190.9-2.2	&		&	23	&	11	\\
G205.5+0.5	&	Monoceros Nebula	&	43	&	15	\\
G260.4-3.4	&	Puppis A, MSH 08-44	&	112	&	39	\\
G292.2-0.5	&		&	89	&	31	\\
G296.5+10.0	&	PKS 1209?51/52	&	64	&	22	\\
G296.7-0.9	&		&	99	&	34	\\
G296.8-0.3	&	1156-62	&	62	&	27	\\
G308.4-1.4	&		&	76	&	39	\\
G315.4-2.3	&	RCW 86, MSH 14-63	&	69	&	30	\\
G327.4+0.4	&	Kes 27	&	87	&	37	\\
G327.6+14.6	&	SN1006	&	76	&	33	\\
G332.4-0.4	&	RCW 103	&	280	&	97	\\
G337.0-0.1	&	CTB 33	&	306	&	132	\\
G337.8-0.1	&	Kes 41	&	224	&	78	\\
G344.7-0.1	&		&	115	&	40	\\
G346.6-0.2	&		&	184	&	64	\\
G349.7+0.2	&		&	629	&	219	\\
G352.7-0.1	&		&	111	&	48	\\
\enddata
\end{deluxetable}


\begin{thebibliography}{}

\bibitem[Adams et al.(2013)]{adams13} Adams, M., Colella, P., Graves, D. T., Johnson, J. N., Keen, N. D. et al., \emph{Chombo Software Package for AMR Applications-Design Document}, 2013,  Lawrence Berkeley National Laboratory Technical Report LBNL-6616E  

\bibitem[Aharonian et al.(2017)]{ahar17} Aharonian, F., Sun, X.-n., \& Yang, R.-z.\ 2017, \aap, 603, A7 

\bibitem[Amato \& Blasi(2009)]{amato09} Amato, E., \& Blasi, P.\ 2009, \mnras, 392, 1591

\bibitem[Arbutina et al.(2012)]{arbo12}Arbutina, B., Uro\v{s}evi\'{c}, D., Andjeli\'{c}, M. M., Pavlovi\'{c}, M. Z., Vukoti{\' c}, B.\ 2012,\apj, 746, 79

\bibitem[Arbutina et al.(2013)]{arbo13} Arbutina, B., Uro\v{s}evi\'{c}, D., Vu{\v c}eti\'{c}, M. M., Pavlovi\'{c}, M. Z., Vukoti{\' c}, B.\ 2013,
\apj, 777, 31

\bibitem[Beck \& Krause(2005)]{bic05} Beck, R., Krause, M.\ 2005,
Astron. Nachr., 326, 414

\bibitem[Beck(2015)]{beck16} Beck, R.\ 2015, \aapr, 24, 4

\bibitem[Bell(1978)]{bell78} Bell, A.~R.\ 1978, \mnras, 182, 443

\bibitem[Bell(2004)]{bell04} Bell, A.~R.\ 2004, \mnras, 353, 550

\bibitem[Berezhko et al.(2002)]{bif02} Berezhko, E.~G., Ksenofontov, L.~T., \& V{\"o}lk, H.~J.\ 2002, \aap, 395, 943

\bibitem[Berezhko \& V{\"o}lk (2004)]{bif04} Berezhko, E.~G., V{\"o}lk, H.~J.\ 2004, \aap, 427, 525

\bibitem[Berezhko \& V{\"o}lk(2006)]{bif06} Berezhko, E.~G., \& V{\"o}lk, H.~J.\ 2006, \aap, 451, 981

\bibitem[Blasi(2002)]{blasi02} Blasi, P., \ 2002, Astropart. Phys., 16, 429


\bibitem[Blasi et al.(2005)]{blasi05} Blasi, P., Gabici, S., \& Vannoni, G.\ 2005, \mnras, 361, 907

\bibitem[Blondin \& Ellison(2001)]{blondin01} Blondin, J.~M., \& Ellison, D.~C.\ 2001, \apj, 560, 244

\bibitem[Borkowski et al.(2010)]{borkow10} Borkowski K. J., Reynolds S. P., Green D. A., Hwang U., Petre R., Krishnamurthy K., Willett R.\ 2010, \apjl, 724, L161

\bibitem[Brugger et al.(2012)]{brugg12} Brugger, C. E. et al., 2012, High Performance Visualization - Enabling Extreme-Scale Scientific Insight: "VisIt: An End-User Tool For Visualizing and Analyzing Very Large Data" (Chapman and Hall/CRC)

\bibitem[Caprioli et al.(2009)]{caprioli09} Caprioli, D., Blasi, P., Amato, E., \& Vietri, M.\ 2009, \mnras, 395, 895

\bibitem[Caprioli(2012)]{capri12} Caprioli, D.\ 2012, \jcap, 7, 038

\bibitem[Caprioli \& Spitkovsky(2014)]{capri14} Caprioli, D., \& Spitkovsky, A.\ 2014, \apj, 794, 46

\bibitem[Duric(1990)]{dju90} Duric, N.\ 1990, in IAU Symp. 140, Galactic and Intergalactic Magnetic Fields, ed. R. Beck, P. P. Kronberg, \& R. Wielebinski (Dordrecht: Kluwer), 235

\bibitem[Duric et al.(1995)]{dju95}
Duric, N., Gordon, S. M., Goss, W. M., Viallefond, F., Lacey, C.\ 1995, \apj, 445, 173

\bibitem[Ellison et al.(2004)]{ellis04} Ellison, D.~C., Decourchelle, A., \& Ballet, J.\ 2004, \aap, 413, 189

\bibitem[Ferrand et al.(2010)]{ferrand10} Ferrand, G., Decourchelle, A., Ballet, J., Teyssier, R., \& Fraschetti,F.\ 2010, \aap, 509, L10

\bibitem[Gosachinskii(2005)]{gosa05} Gosachinskii, I.~V.\ 2005, Astronomy Letters, 31, 179 

\bibitem[Green et al.(2008)]{green08} Green, D.~A., Reynolds, S.~P., Borkowski, K.~J., et al.\ 2008, \mnras, 387, L54

\bibitem[Han(2017)]{han17} Han, J. L.\ 2017, \araa, 55, 111

\bibitem[Helder et al.(2012)]{helder12} Helder, E.A., Vink, J., Bykov A.M., Ohira, Y., Raymond, J.C., Terrier, R.\ 2012, Space Sci. Rev., 173, 369

\bibitem[Kang et al.(2013)]{kang13} Kang, H., Jones, T.~W., \& Edmon, P.~P.\ 2013, \apj, 777, 25

\bibitem[Kosenko et al.(2014)]{kosenko14} Kosenko, D., Ferrand, G., \& Decourchelle, A.\ 2014, \mnras, 443, 1390

\bibitem[Kothes et al.(2006)]{kothes06} Kothes, R., Fedotov, K., Foster, T.~J., \& Uyan{\i}ker, B.\ 2006, \aap, 457, 1081

\bibitem[Lazendic \& Slane(2006)]{laze06} Lazendic, J.~S., \& Slane, P.~O.\ 2006, \apj, 647, 350

\bibitem[Lee at al.(2012)]{lee12} Lee, S.-H., Ellison, D.~C., \& Nagataki, S.\ 2012 \apj, 750, 156

\bibitem[Lequeux(2005)]{leq05} Lequeux, J.\ 2005, The Interstellar Medium (Berlin Heidelberg: Springer-Verlag)

\bibitem[Li et al.(2017)]{li17} Li, H.-B., Jiang, H., Fan, X., Gu, Q., \& Zhang, Y.\ 2017, Nature Astronomy, 1, 0158

\bibitem[Longair(1994)]{long94} Longair, M. S.\ 1994, High Energy Astrophysics Vol. 2, (Cambridge: Cambridge Univ. Press)

\bibitem[Mignone et al.(2007)]{mig07} Mignone, A., Bodo, G., Massaglia, S., Matsakos, T., Tesileanu, O., Zanni, C., Ferrari, A.\ 2007, \apjs, 170, 228

\bibitem[Mignone et al.(2012)]{mig12} Mignone, A., Zanni, C., Tzeferacos, P., van Straalen, B., Colella, P., Bodo, G.\ 2012, \apjs, 198, 7

\bibitem[Morlino et al.(2009)]{morlino09} Morlino, G., Amato, E., \& Blasi, P.\ 2009, \mnras, 392, 240

\bibitem[Morlino \& Caprioli(2012)]{morlino12} Morlino, G., \& Caprioli, D.\ 2012, \aap, 538, A81

\bibitem[Orlando et al.(2012)]{orlando12} Orlando, S., Bocchino, F., Miceli, M., Petruk, O., \& Pumo, M.~L.\ 2012, \apj, 749, 156

\bibitem[Pacholczyk(1970)]{pach70} Pacholczyk, A. G.\ 1970, Radio Astrophysics (San Francisco, CA:Freeman)

\bibitem[Pavlovi{\' c} et al.(2014)]{pav14} Pavlovi{\' c}, M.~Z., Dobardzic, A., Vukotic, B., \& Urosevic, D.\ 2014, Serbian Astronomical Journal, 189, 25

\bibitem[Pavlovi{\' c}(2017)]{pav17} Pavlovi{\' c}, M.Z.\ 2017, \mnras, 468, 1616

\bibitem[Pfrommer et al.(2017)]{pfro17} Pfrommer, C., Pakmor, R., Schaal, K., Simpson, C.~M., \& Springel, V.\ 2017, \mnras, 465, 4500

\bibitem[Ptuskin (2007)]{ptuskin07} Ptuskin, V.S.\ 2007, Phys. Usp., 50, 534

\bibitem[Reynolds et al.(2008)]{reyn08} Reynolds, S. P., Borkowski, K. J., Green, D. A., Hwang, U., Harrus, I., Petre, R., 2008, \apjl, 680, L41 

\bibitem[Reynolds (2008)]{reynolds08} Reynolds, S.~P.\ 2008, \araa, 46, 89

\bibitem[Salvesen, Raymond \& Edgar (2009)]{sre09} Salvesen, G, Raymond, J.~C., \& Edgar, R.~J.\ 2009, \apj, 702, 327

\bibitem[Sarbadhicary et al.(2017)]{sarb17} Sarbadhicary, S.~K., Badenes, C., Chomiuk, L., Caprioli, D., \& Huizenga, D.\ 2017, \mnras, 464, 2326

\bibitem[Truelove \& McKee(1999)]{tmc99} Truelove, J.~K., \& McKee, C.~F.\ 1999, \apjs, 120, 299

\bibitem[Uro{\v s}evi{\' c}(2014)]{uro14} Uro{\v s}evi{\' c}, D.\ 2014,
Ap. S. Sci., 354, 541

\bibitem[V{\"o}lk et al.(2005)]{volk05} V{\"o}lk, H.~J., Berezhko, E.~G., \& Ksenofontov, L.~T.\ 2005, \aap, 433, 229

\bibitem[Yang et al.(2014)]{yang14} Yang, R-z., Zhang, X., Yuan, Q., \& Liu, S.\ 2014, \aap, 567, 23

\bibitem[Yoast-Hull et al.(2016)]{yoast16} Yoast-Hull, T.~M., Gallagher, J.~S., \& Zweibel, E.~G.\ 2016, \mnras, 457, L29


\bibitem[Yuan et al.(2012)]{yuan12} Yuan, Q., Liu, S., \& Bi, X.\ 2012, \apj, 761, 133

\bibitem[Zhou et al.(2016)]{zhou16} Zhou, X., Yang, J., Fang, M., et al.\ 2016, \apj, 833, 4

\bibitem[Zirakashvili \& Aharonian(2007)]{zirak07} Zirakashvili, V.~N., \& Aharonian, F.\ 2007, \aap, 465, 695

\bibitem[Zoglauer et al.(2015)]{zog15} Zoglauer, A., Reynolds, S.~P., An, H., et al.\ 2015, \apj, 798, 98 


\end{thebibliography}
\end{document}